\documentclass{bioinfo}
\copyrightyear{2012}
\pubyear{2012}

\usepackage{xspace}
\newcommand{\up}{UniProtKB\xspace}
\newcommand{\swp}{Swiss-Prot\xspace}
\newcommand{\tr}{TrEMBL\xspace}
\newcommand{\pw}{power-law\xspace}

\usepackage{microtype}

\usepackage{booktabs}

\begin{document}
\firstpage{1}

\title[An approach to describe and analyse bulk annotation quality]{An
  approach to describing and analysing bulk biological annotation
  quality: A case study using UniProtKB} \author[M. J. Bell
  \textit{et~al}]{Michael J. Bell\,$^{1}$, Colin S. Gillespie\,$^{2}$,
  Daniel Swan\,$^{3, \footnote{Current affiliation: Oxford Gene
      Technology, Yarnton, Oxfordshire, UK.}}$, Phillip
  Lord\,$^{1,}$\footnote{to whom correspondence should be addressed}}

\address{$^{1}$School of Computing Science, Newcastle University, Newcastle-Upon-Tyne, NE1 7RU, UK.\\
$^{2}$School of Mathematics \& Statistics, Newcastle University, Newcastle-Upon-Tyne, NE1 7RU, UK. \\
$^{3}$Bioinformatics Support Unit, ICAMB, Medical School, Newcastle University, Newcastle-Upon-Tyne, NE1 7RU, UK.}

\maketitle

\begin{abstract}
\section{Motivation:}

Annotations are a key feature of many biological databases, used to
convey our knowledge of a sequence to the reader. Ideally, annotations
are curated manually, however manual curation is costly, time
consuming and requires expert knowledge and training. Given these
issues and the exponential increase of data, many databases implement
automated annotation pipelines in an attempt to avoid un-annotated
entries. Both manual and automated annotations vary in quality between
databases and annotators, making assessment of annotation reliability
problematic for users. The community lacks a generic measure for
determining annotation quality and correctness, which we look at
addressing within this paper. Specifically we investigate word reuse
within bulk textual annotations and relate this to Zipf's Principle of
Least Effort. We use \up as a case study to demonstrate this approach
since it allows us to compare annotation change, both over time and
between automated and manually-curated annotations.

\section{Results:}

By applying \pw distributions to word reuse in annotation, we show
clear trends in \up over time, which are consistent with existing
studies of quality on free text English. Further, we show a clear
distinction between manual and automated analysis and investigate
cohorts of protein records as they mature. These results suggest that
this approach holds distinct promise as a mechanism for judging
annotation quality.

\section{Availability:}

Source code and supplementary data are available at the authors
website: \href{http://homepages.cs.ncl.ac.uk/m.j.bell1/annotation/}
{http://homepages.cs.ncl.ac.uk/m.j.bell1/annotation/}

\section{Contact:} \href{phillip.lord@newcastle.ac.uk} {phillip.lord@newcastle.ac.uk}
\end{abstract}

\section{Introduction}

A key descriptive feature of biological data is its \emph{annotation}:
a textual representation of the biology associated with the
data. Biologists use these annotations to understand and contextualise
data in biological sequence databases. Annotations play an essential
role in describing and developing the users' knowledge of a given
sequence and can form the foundation for further
research~\citep{Jones07Estimating}. Some annotation is structured, for
example, using an ontology~\citep{Stevens09Application} or keyword
list. However, free text annotation often contains the richest
biological knowledge~\citep{Camon05Evaluation}, but while free text is
appropriate for human comprehension it is difficult to interpret
computationally.

The current `gold standard' for annotation is a set of reviewed and
manually-curated entries~\citep{Curwen04Ensembl}. However,
manually-curated annotation is labour-intensive, time consuming and
costly. To cope with the amount of data, which is typically increasing
exponentially, many resources and projects generate annotations
computationally~\citep{Uniprot03}. Automated annotations are more
prone to errors than their manual
counterparts~\citep{Gilks02Modeling}, with several studies suggesting
high levels of misannotation in automated
annotation~\citep{Schnoes09Error, Andorf07Exploring,
  Jones07Estimating}. It can be hard, even impossible, to determine
the source from which an error has propagated~\citep{Buza08Gene}
causing significant problems for biologists. Annotation quality is
not consistent across all databases and
annotators~\citep{Dolan05Procedure}, whether curated manually or
automatically. It can, therefore, be difficult to determine the level
of quality, maturity or correctness of a given textual
annotation. However users often incorrectly assume that annotations
are of consistent quality and correctness~\citep{Ussery04Genome}.

Currently there are few standard metrics for assessing annotation
quality.  Annotations are frequently assigned a score, using a variety
of methods. These approaches include assigning confidence scores to
annotations based on their stability~\citep{Gross09Estimating} or
combining the breadth (coverage of gene product) and the depth (level
of detail) for the terms in the Gene Ontology
(GO)~\citep{Buza08Gene}. However, while deeper nodes within an
ontology are generally more specialised, these measures are
problematic; first GO has three root domains and second an ontology,
such as GO, is a graph not a tree, therefore depth is not necessarily
meaningful. Other methods~\citep{Rogers09Evidence, Buza08Gene,
  Pal05Inference} use evidence codes as a basis for an annotations
reliability, although ironically, the Gene Ontology annotation manual
explicitly states that evidence codes should NOT be used in this
way~\citep{GOGuide}, describing rather the type of evidence not its
strength.

All of these approaches rely upon additional information to determine
annotation quality. Resources, such as sequence databases, vary in
their structures and in the additional information stored. For
example, not all resources use evidence codes and these codes are not
comparable between resources~\citep{Lord03Investigating}; likewise, it
is not generally possible to use methods based on an ontological
hierarchy for non-ontological resources.

Most resources carry some annotations which are unstructured, free
text. Therefore, a quality metric that can be derived purely from
textual annotation would potentially allow any resource to be
analysed and scored.  There are various measures for analysing the
quality of text, such as the Flesch-Kincaid Readability
Test~\citep{Flesch48Readability} and SMOG
Grading~\citep{McLaughlin69SMOG}. These metrics are generally based
around readability, or reading-age; that is the literary quality of
the text, rather than correctness and quality of the subject matter.

In this paper, we report on a bulk analysis of textual annotation in
\up, attempting to understand whether we can exploit our knowledge of
changes in the annotation over time as a mechanism for developing a
quality measure of biological correctness. We investigate word
occurrences, and their changes over time, as reflected in their
distribution; we show that using these relationships we are able to
detect large-scale changes in the annotation process; and we
demonstrate that the parameters of these relationships also
change. Specifically, we fit a \pw distribution to the extracted word
occurrences and extract a value, called $\alpha$. We relate this
$\alpha$ value to Zipf's principle of least effort, which states it is
human nature to take the path of least effort to achieve a
goal. Broadly, higher values of $\alpha$ indicate a resource which is
easier for the reader, while lower values are easier for the
annotator.

While manually curated annotation is generally accepted as the most
accurate~\citep{Curwen04Ensembl}, a significant problem is the lack of more
explicit gold standard data sets~\citep{James11Multiple, Roberts10COMBREX}.
This makes defining a quality measure somewhat troublesome. Our investigation
into whether changes in word distribution are representative of quality and
maturity in the annotation show that these forms of measures can detect
large-scale features of annotation, and that clear trends appear as \up
matures and grows over time. These trends are often reflective of our
\textit{a priori} judgements of quality within \up, for example, a distinction
between manual and automated annotation. In the absence of a gold standard, we
believe that this represents reasonable evidence that this form of analysis
may be used as the basis for a quality metric for textual annotation.

\begin{methods}
\section{Methods}

\subsection{Data extraction from UniProtKB}

The UniProt KnowledgeBase (\up)~\citep{Uniprot10} consists of two sections:
\up/\swp, which is reviewed and manually annotated, and \up/\tr which is
unreviewed and automatically annotated. The first version of \swp was released
in 1986, with \tr appearing in 1996. Releases of \tr were initially more
frequent than \swp, meaning the databases were released independently. In
Table~\ref{tab:relDates}, we map between each \swp release and the nearest
version of \tr. This allows us to compare the quality of manually and
automatically curated annotation at similar points in time.

Following the formation of the UniProt Consortium in 2002, the releases of the
two databases were synchronised (from 2004). The correct names are now
technically \up /\swp and \up /\tr. We will use the following naming approach
for clarity:

\begin{itemize}
  \item {\bf{UniProt}} -- Refers to the UniProt Consortium.
  \item {\bf \swp} -- Refers to \swp entries prior to the formation of
    the UniProt Consortium.
  \item {\bf \tr} -- Refers to \tr entries prior to the formation of
    the UniProt Consortium.
  \item {\bf \up} -- Refers to the combination of both \swp and \tr
    datasets. 
\end {itemize}

Where necessary we will explicitly write \up/\swp or \up/\tr. This naming
scheme allows us to refer to post-\up versions of \up/\swp and \up/\tr with
the same number, starting from version two of \up\footnote{Version two was the
  first major release containing \swp version 44 and \tr version 27.}. We can
investigate annotation change over time, as complete datasets
for historical versions of \up and \swp are made available by UniProt on their
FTP server, with the exception of \swp versions 1-8 and 10 which were never
archived. Pre-\up/\tr releases were kindly made available to us by UniProt.

\begin{table}[!t]
\caption{\textbf{Mapping between \tr and \swp release dates.} For each
  version of \swp, we have associated the nearest version of \tr based
  on release date.}
\footnotesize
\begin{tabular}{@{}llll@{}}
\toprule
Date   & \swp version & Date & \tr version \\
\midrule
Oct-96 & 34 & Nov-96 & 1 \\
Nov-97 & 35 & Jan-98 & 5 \\
Jul-98 & 36 & Aug-98 & 7 \\
Dec-98 & 37 & Jan-99 & 9 \\
Jul-99 & 38 & Aug-99 & 11 \\
May-00 & 39 & May-00 & 13 \\
Oct-01 & 40 & Oct-01 & 18 \\
Feb-03 & 41 & Mar-03 & 23 \\
Oct-03 & 42 & Oct-03 & 25 \\
Mar-04 & 43 & Mar-04 & 26 \\
\bottomrule
\end{tabular}
\label{tab:relDates}
\end{table}

Our extraction approach, also summarised in
Figure~\ref{fig:extractionprocess}, involves four key steps:

\begin{enumerate}
\item The UniProt FTP
  server\footnote{\href{ftp.uniprot.org/pub/databases/uniprot/}{ftp.uniprot.org/pub/databases/uniprot/}}
  provides complete datasets for past versions of \swp and \up in flat
  file format.
\item \up flat files adhere to a strict structure, as detailed in the
  \up user manual~\citep{UPManual}. A Java framework was created that
  allowed \up comment lines to be correctly extracted.
\item Over time annotations in \up have become more structured with the
  addition of topic headings (e\.g\. ``subcellular location'' and
  ``function''). These headings are ignored by our analysis. We also remove
  punctuation, the `CC' identifier, brackets and whitespace.
\item The final step in this process is to output a list of all words and
  their frequency for all annotations in a given database version.
\end{enumerate}

In order to ensure accurate data extraction, we checked that
the number of entries parsed (Figure~\ref{fig:extractionprocess}, step
2) matched the number expected from the release notes. Additionally the
list of headings (comment blocks and properties) removed
(Figure~\ref{fig:extractionprocess}, step 3) were noted, along with
their frequency, to ensure only headings described in the \up manual
were removed.  Finally, a random selection of records were manually
checked against parsed outputs (Figure~\ref{fig:extractionprocess},
all steps).

\begin{figure}[!tpb]
  \centering
  \includegraphics[width=0.40\textwidth]{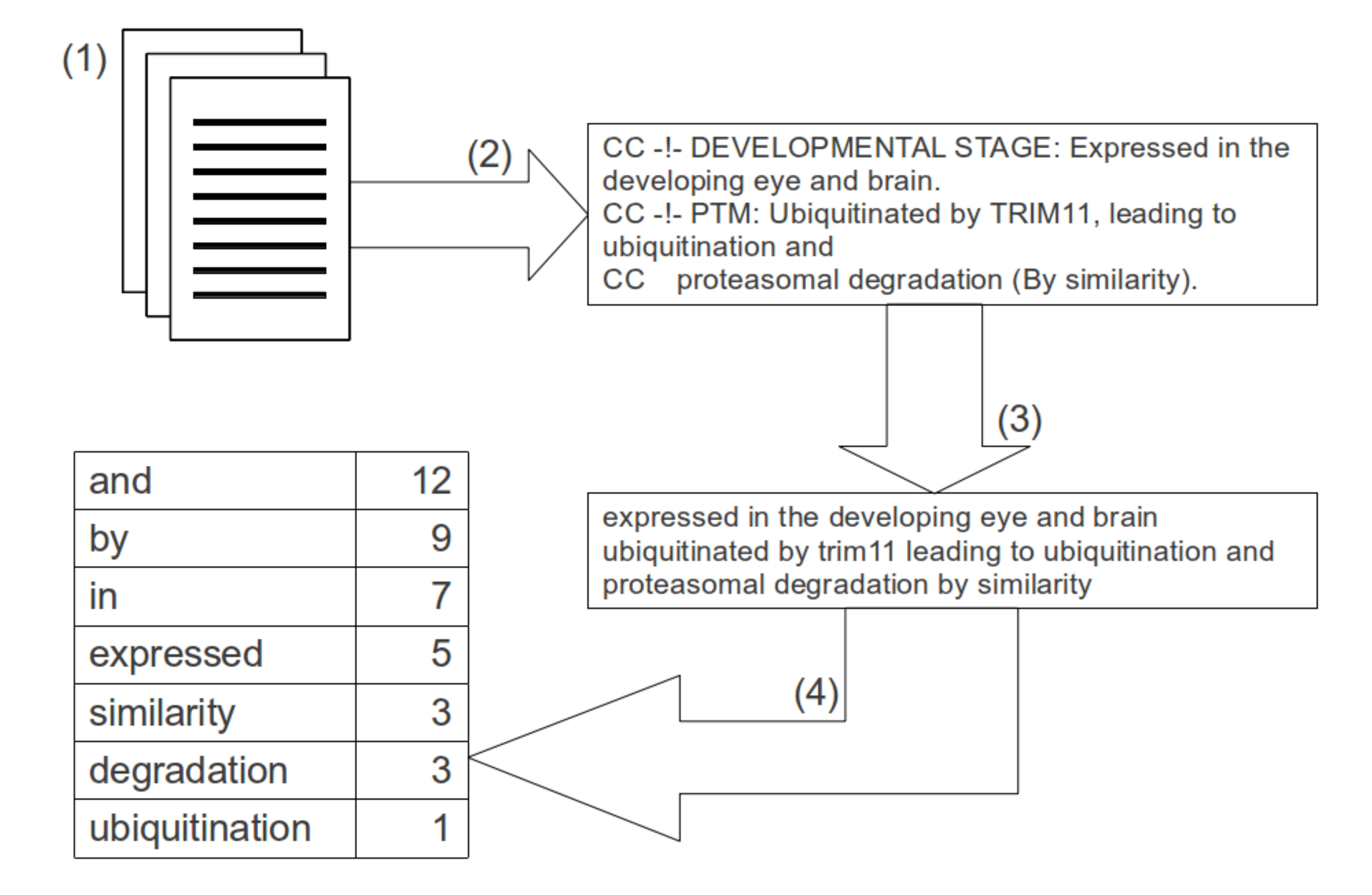}
  \caption{ \textbf{Outline view of the data extraction process.} (1)
    Initially we download a complete dataset for a given database
    version in flat file format. (2) We then extract the comment lines
    (lines beginning with `CC', the comment indicator). (3) We remove
    comment blocks and properties (as defined in the \up manual~\citep{UPManual}),
    punctuation, `CC', brackets and make words lower case, so as to
    treat them as case insensitive. (4) Finally, we count the
    individual words and update the occurrence of each word total
    count.}
  \label{fig:extractionprocess}
\end{figure}

\subsection{Model fitting}
\label{model_fitting}

When developing a framework to model the occurrence of words, a variety of
competing models were considered. These ranged from relatively simple
distributions, such as the exponential and log-normal, to more complex mixture
models. However, the \pw distribution achieved a good balance between model
parsimony and fit.

In this paper, we only deal with the discrete \pw distribution (see
\citep{Clauset09Powerlaw} for a discussion on power-laws). The discrete \pw
distribution has probability mass function
\[
p(x) = \frac{x^{-\alpha}}{\zeta(\alpha, x_{\min})}
\]
where
\[
\zeta(\alpha, x_{min}) = \sum_{n=0}^{\infty} (n + x_{\min})^{-\alpha}
\]
is the generalised or Hurwitz zeta function. 

To fit the \pw distribution, we followed the Bayesian paradigm. We
assumed a proper uniform $U(1, 5)$ prior for $\alpha$. Since the
posterior distribution for the parameters is analytically intractable,
we integrate out the uncertainty using a MCMC algorithm. The parameter
space was explored using a Gaussian random walk. The Markov chain
reached equilibrium very quickly and only a small amount of thinning
was necessary.

We modelled multiple datasets simultaneously using a ``fixed-effects''
approach.  Let $i$ denote the dataset of interest, then we aim to
infer the parameter
\[
\alpha_i = \alpha + \mu_i
\]
where $\alpha$ is the coefficient for a baseline dataset and $\mu_i$
is difference from this baseline. For example in
Figure~\ref{fig:tremblvsswissprot_alphas}, the baseline dataset is
\up/\swp version 16 and $\mu_i$ represents the change in the $\alpha$
coefficient from \up/\swp version 16.

Throughout this paper, we set $x_{\min} = 50$, which we determined
using the BIC criteria when fitting all the datasets in
Figure~\ref{fig:tremblvsswissprot_alphas}. However, the conclusions
are not sensitive to changes of $x_{\min}$. For smaller values of
$x_{\min}$, the credible region is reduced since there is more data,
conversely increasing $x_{\min}$ to around 200 increases the credible
regions slightly.

The fitting of a \pw corresponds to the exponent of the regression
line represented by $\alpha$. Given that the graph and $\alpha$ value
is based on the underlying text, it is plausible that the $\alpha$
value could provide a measurement to assess the underlying textual
quality. Indeed, it has been previously
suggested~\citep{Ferrer05Variation} that the $\alpha$ value is related
to Zipf's principle of least
effort~\citep{Zipf1949Human}\footnote{Additionally, Zipf has shown
  that a words occurrence is inversely proportional to its rank
  (Zipf's Law). Zipf's law, Pareto's law and \pw distributions are all
  types of \pw that are, essentially, different ways of looking at the
  same thing~\citep{Adamic02Zipf}.}. This principle states that it is
human nature to take the path of least effort when achieving a
goal. For example, an annotator can create an annotation with generic
terms (least effort for the annotator, more work for the reader) or
with precise and specialist terms (least effort for the reader, more
work for the annotator). Table~\ref{tab:alphas} shows $\alpha$ values
that have been extracted from a variety of texts, that give confidence
to this claim. We can use this information as a basis for quality;
texts which require minimal effort for the reader, due to expertly
curated annotation, are deemed to be of high quality.

\begin{table*}[!t]
  \centering
  \caption{\textbf{Relationship between $\boldsymbol\alpha$ value and
      Zipf's principle of least effort.} For $\alpha$ values less than
    1.6 or greater than 2.4, we have no corresponding effort level as
    the text is treated as incomprehensible.}
  \begin{tabular}{@{}lp{12.5cm}l@{}}
    \toprule \textbf{$\boldsymbol\alpha$ value} & \textbf{Examples in
      literature} & \textbf{Least effort for} \\ 
    \midrule 
    $\alpha < 1.6$
    & Advanced schizophrenia~\citep{Zipf1949Human,
      Piotrowska04Statistical}, young
    children~\citep{Piotrowska04Statistical, Brillouin04Science} & -  \\ 
    $1.6 \leq \alpha < 2$ & Military combat texts~\citep{Piotrowska04Statistical},
    Wikipedia~\citep{Serrano09Modeling}, Web pages listed on the open
    directory project~\citep{Serrano09Modeling} & Annotator \\ 
    $\alpha = 2$ & Single author texts~\citep{Balasub96Quant} & Equal effort levels \\ 
    $2 < \alpha \leq 2.4$ & Multi author texts~\citep{Ferrer05Decoding} & Audience \\ 
    $\alpha > 2.4$ &  Fragmented discourse schizophrenia~\citep{Piotrowska04Statistical} &
    - \\ \bottomrule
  \end{tabular}
  \label{tab:alphas}
\end{table*}

\end{methods}

\section{Results}

\subsection{Does annotation in \up obey a \pw distribution?}

\begin{figure*}[!t]
  \centering
  \includegraphics[width=0.39\textwidth]{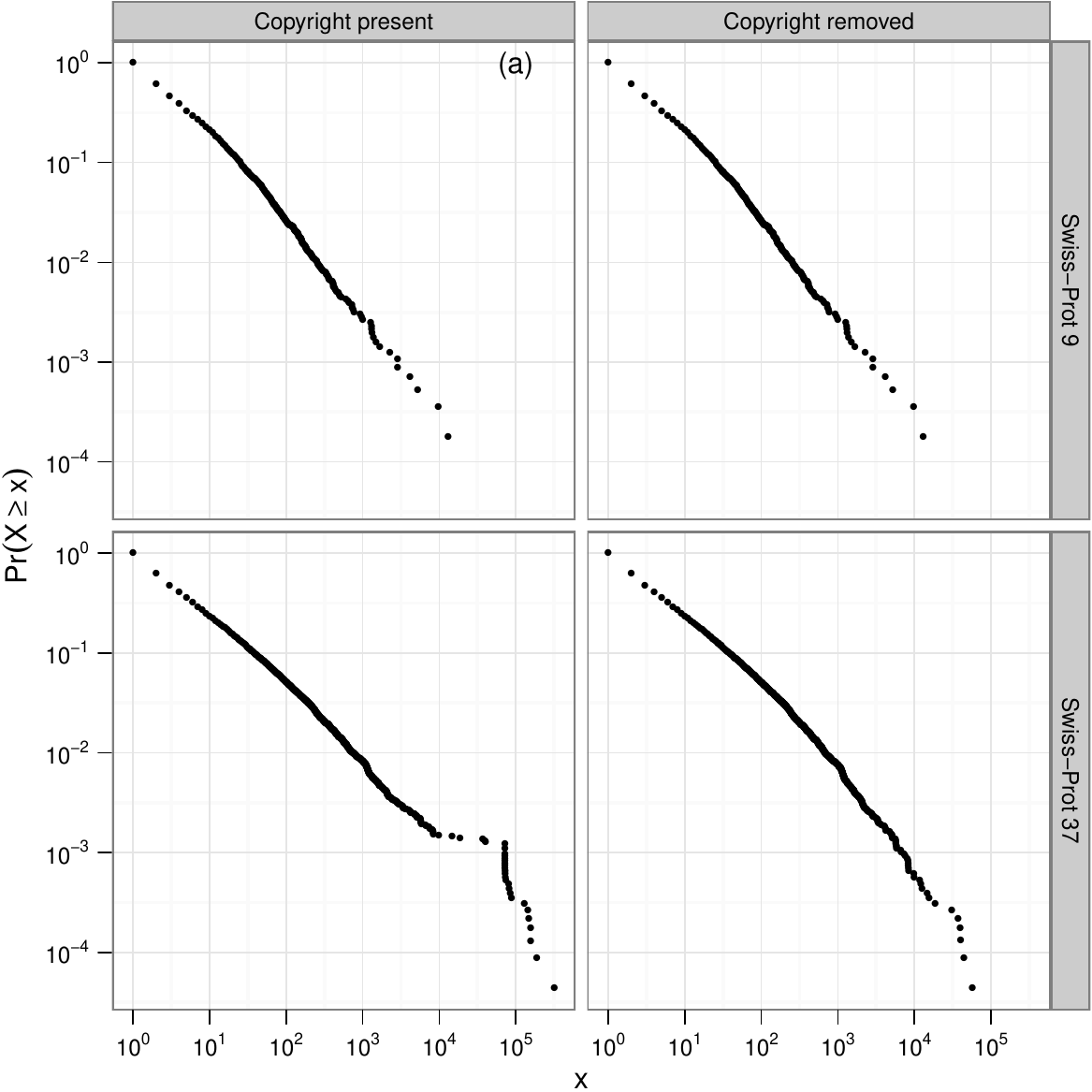}
  \includegraphics[width=0.39\textwidth]{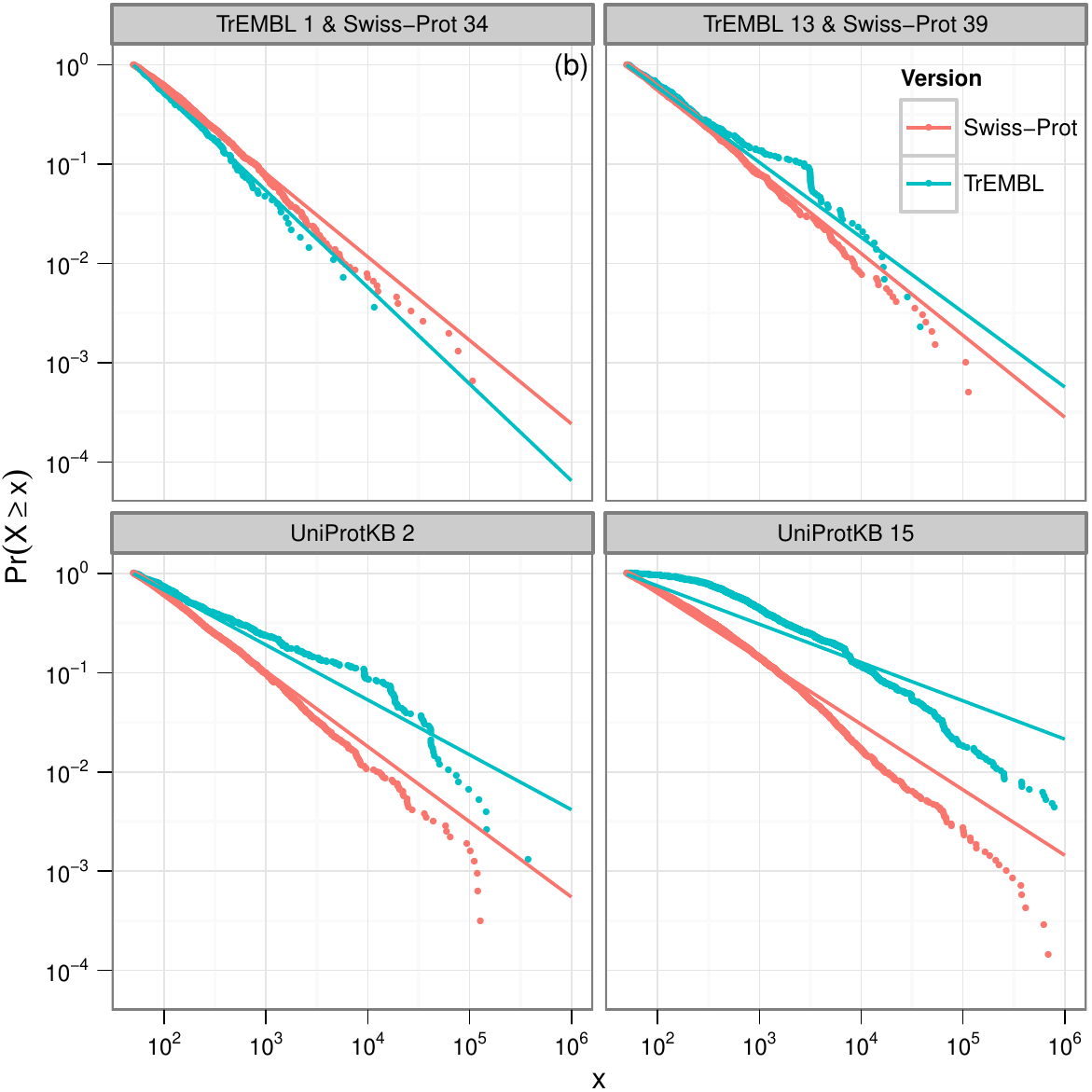}

  \caption{Cumulative distributions of words for various \swp and \tr
    versions, shown with logarithmic scales. The size (number of words) is
    shown along the $X$ axis while the probability is shown on the $Y$ axis. A
    point on the graph represents the probability that a word will occur $x$
    or more times. For example, the upper left most point represents the
    probability of 1 (i.e. $10^{0}$) that a given word will occur once (i.e.
    $10^{0}$) or more times. A word must occur at least once to be included.
    Words occurring very frequently are presented in the bottom right of the
    graph. (a) Shows the resulting graphs for \swp version 9 (November 1988)
    and \swp version 37 (December 1998), with and without copyright. The
    distinct structure visible between $x=10^4$ and $x=10^5$ in \swp version
    37 (bottom left panel) is caused by the copyright statement declaration.
    \swp version 9 operates as a control to show that the attempted removal of
    copyright has no effect where no copyright information is present. (b)
    Shows the data with fitted \pw distributions for an even subset of
    historical versions of \swp and the co-ordinate release of \tr.}

  \label{fig:zipfs_copyright}  
\end{figure*}

Power-laws have been shown to exist in numerous man-made and natural
phenomena~\citep{Clauset09Powerlaw}. The link between Zipf's principle
of least effort and $\alpha$ was originally based on natural
language. If a \pw distribution is a measure of quality, we would
expect that a \pw distribution is more likely to occur in
human-curated annotation rather than annotations produced
automatically. Therefore, for our initial analysis, we selected \swp
as a gold standard resource. Results are shown in
Figure~\ref{fig:zipfs_copyright}(a), for two versions of \swp. We can
see that annotation does broadly obey a \pw, although with a distinct
``kink'' in \swp version 37, between $x=10^4$ and $x=10^5$.

Inspection of the words in this region showed that this structure is
artifactual, resulting not from annotation \emph{per se} but from
copyright and license information, which are included in the CC lines,
along with biological information. These copyright statements were
first introduced into \swp at version 37, with wording changes at \up
versions 4 and 7. From this analysis, we show that we can detect the
introduction of a large amount of material with no biological
significance into the annotation. This demonstrates that the \pw can
be used as a partial measure of quality, albeit for detecting
artifacts.

For subsequent analysis we removed the copyright statements. Updated graphs,
with copyright statements removed, are also shown in
Figure~\ref{fig:zipfs_copyright}(a). Inspection of these graphs show that the
slope for the head and tail increase at different rates with \swp versions.
This is a result of a marked two-slope behaviour which is commonly seen for
mature resources, such as large complex natural languages~\citep{Ha06Zipf,
  Cancho01Regimes}. These graphs follow a \pw distribution reasonably well.
However Figure~\ref{fig:zipfs_copyright}(b) shows some versions of \tr where a
\pw does not fit that well. As discussed in Section~\ref{model_fitting},
various competing models were considered, with a \pw being chosen partly due
to the simplicity of its output, $\alpha$. It is clear that this change over
time requires further analysis.

\subsection{How do the distributions change over time?}

Although it is useful for demonstrating the two-slope behaviour, this view
makes it difficult to see change over time. Given that the main analytical
value comes from the extracted $\alpha$ values, subsequent graphs show just
these values for different database versions. This approach
allows us to investigate the change over time by looking at all
historical data simultaneously. The resulting graphs from this analysis is
shown in the top half of Figure~\ref{fig:tremblvsswissprot_alphas}.

As Figure~\ref{fig:tremblvsswissprot_alphas} shows, the annotation in \swp is
changing in its nature over time. $\alpha$ decreases over time for \swp,
suggesting that \swp is becoming optimized toward least effort for the
annotator, rather than the reader. This fits with previous research from
\citeauthor{Manual07Baumgartner}~(\citeyear{Manual07Baumgartner}) suggesting
that the enormous increase in proteins requiring annotation is outstripping
the provision of this annotation. This issue has been acknowledged by UniProt,
with their introduction of automated annotation; therefore, we next
investigate this form of annotation.

\subsection{How does manual annotation compare to automated annotation?}

\begin{figure}[!tpb] 
\centering
\includegraphics[width=0.40\textwidth]{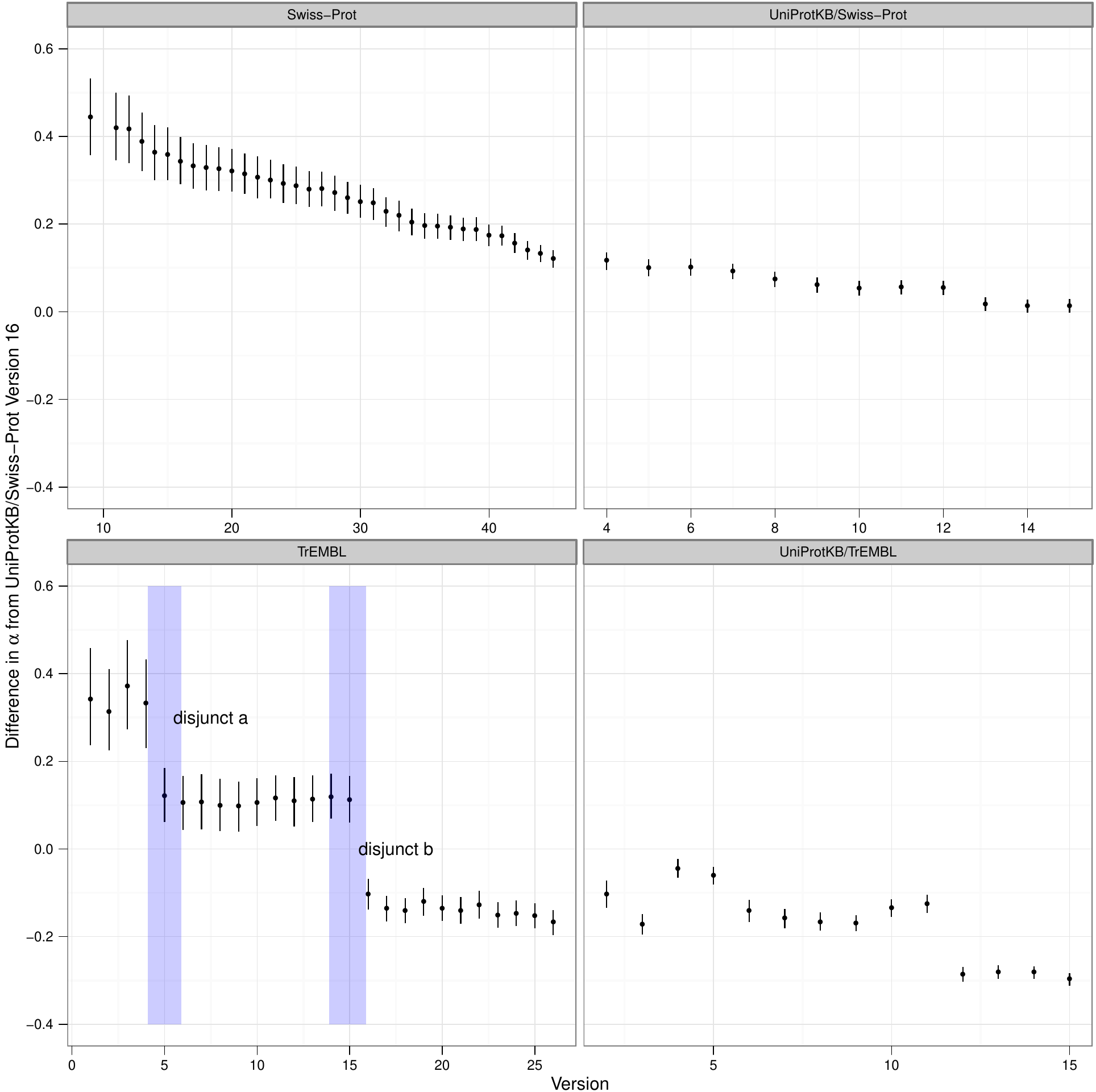}
\caption{$\alpha$ values over time, for each version of \swp and \tr. The
  graph shows the difference in $\alpha$ value (with 95\% credible region)
  from \up/\swp version 16, for which the $\alpha$ value was 1.62. So, for
  example, \swp version 9 has a difference of, approximately, 0.45. Therefore
  the resulting $\alpha$ for \swp version 9 is around 2.07.}
\label{fig:tremblvsswissprot_alphas}
\end{figure}

Within \up proteins are initially annotated automatically and placed into \tr.
Eventually they are manually annotated and placed into \swp. Therefore, \tr
and \swp are ideal resources to compare equivalent human and automated
annotations. Here, we compare these two resources, investigating their
behaviour over time, at equivalent points in time. As previously described,
prior to \up version two, \tr and \swp releases were not synchronized, so we
use the version of \tr released closest in time to each version of \swp, as
show in Table~\ref{tab:relDates}. An evenly spaced subset of these analyses
are shown in Figure~\ref{fig:zipfs_copyright}(b), with
Figure~\ref{fig:tremblvsswissprot_alphas} showing the $\alpha$ values for all
versions of \tr and \swp.

In Figure~\ref{fig:zipfs_copyright}(b), \tr and \swp appear to diverge over
time with \swp demonstrating the behaviour typical of a more mature resource.
\tr shows less maturity, with many words occurring with a high frequency. In
short, \swp appears to show a richer use of vocabulary. We cannot, however,
rule out the possibility that this difference occurs as they are annotating
different proteins. Unfortunately, it is not possible to check a proteins
annotation in both \swp and \tr at the same point in time; once a record is
migrated to \swp, it is removed from subsequent versions of \tr. This is
necessary as \swp is used as a basis for annotation in \tr, so proteins not
removed from \tr would have their automated annotation based on their manual
annotation in \swp. However, the rapid increase in size of both resources,
argues against this explanation.

While \swp shows a relatively regular progression, \tr does not. There are two
significant disjuncts in the relationship where large jumps occur between
releases, as highlighted in Figure~\ref{fig:tremblvsswissprot_alphas}. We also
note a significant rise of total words between these versions, compared to
those for nearby releases (data not shown).

We have identified two plausible explanations for these disjuncts,
based on historical events. Firstly in 1998 (highlighted by
\emph{disjunct a}) a number of new procedures appear to have been
introduced~\citep{Bairoch98SwissProt}.  These approaches include
making use of the ENZYME database, specialised genomic databases and
scanning for PROSITE patterns compatible with an entries taxonomic
range. PROSITE patterns are used to enhance the content of the
comment lines by adding information such as protein function and
subcellular location. Interestingly, prior to this disjunct, the first
four versions of \tr have an $\alpha$ value higher than their \swp
counterpart.

Secondly in 2000, the introduction and development of annotation rules was
planned in \tr which could explain the second jump (highlighted by
\emph{disjunct b})~\citep{Bairoch00SwissProtProtein}. Both of these disjuncts
would be expected to produce an increase in the total amount of annotation, as
well as introducing new words and phrases which would affect the measures
described here. Given the lack of detailed statistics and the age of the
database at this time, UniProt could not confirm these explanations. They did
acknowledge that extensive work in 2001 and early 2002 was carried out to
improve the data, although they believe the scanning of PROSITE was in effect
from \tr version 1.

The increase of total words noted earlier correlates with the increase of
entries into \up; the rate of data being added is exponential. Given this
increase, we find ourselves analysing entries and annotations of mixed age.
Here we have seen the apparent decreasing of quality for complete datasets, of
both \swp and \tr, over time. Following on from this, we wish to explore the
quality of annotations within a set of mature entries.

\begin{figure}[!tpb] 
\centering
  \includegraphics[width=0.7\columnwidth]{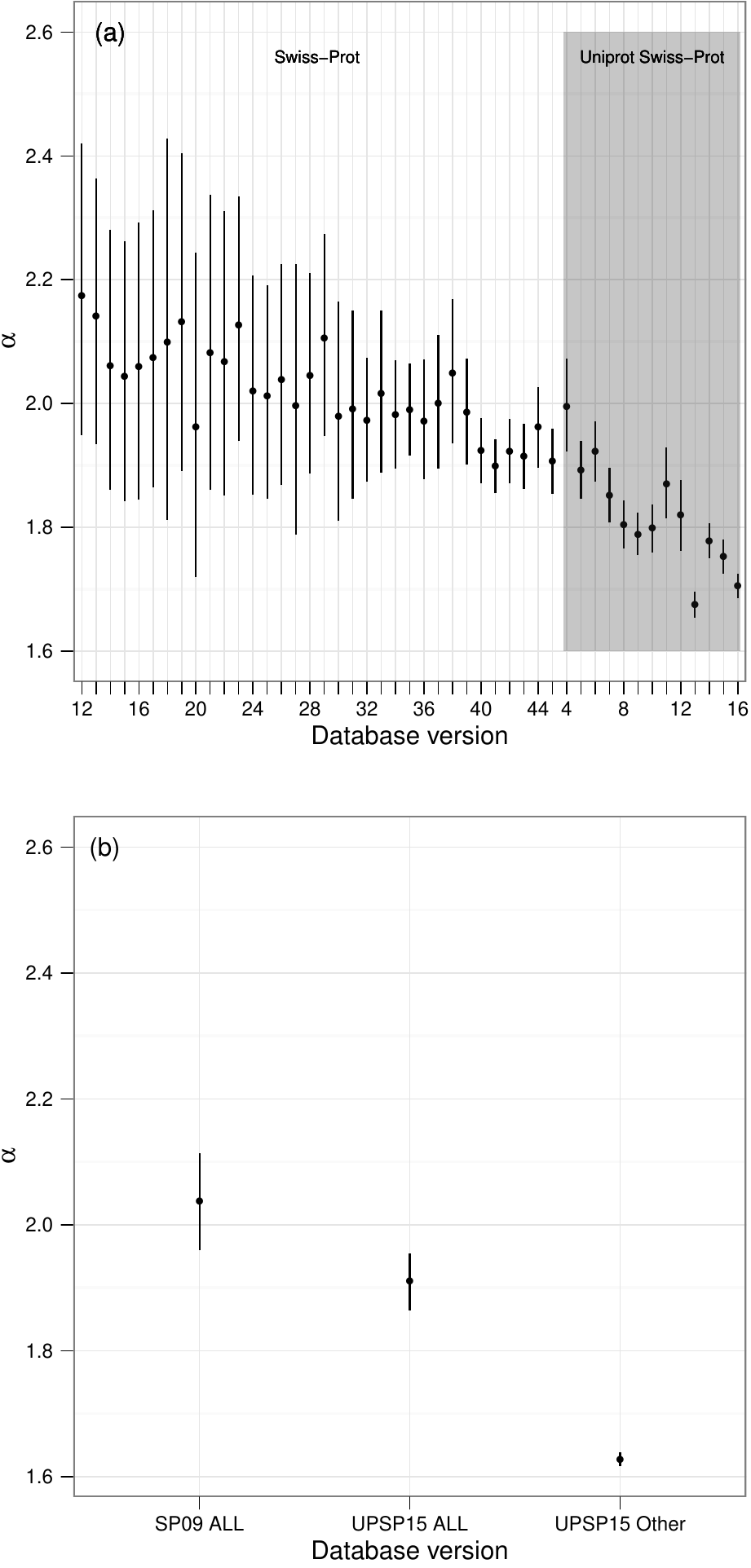}
  \caption{(a) Analysis of those entries that are new to a particular version of
    \swp. (b) $\alpha$ value (with 95\% credible region) for all entries in \swp
    version 9 that are in \up version 15, all entries in \up version 15 that are
    in \swp version 9, and all those in \up version 15, but not in \swp version
    9.}
\label{fig:maturealphas}
\end{figure}

\subsection{Analysing maturity of entries over time and the impact of new annotations}

\begin{figure*}[!tpb]
 \centering \includegraphics[width=0.8\textwidth]{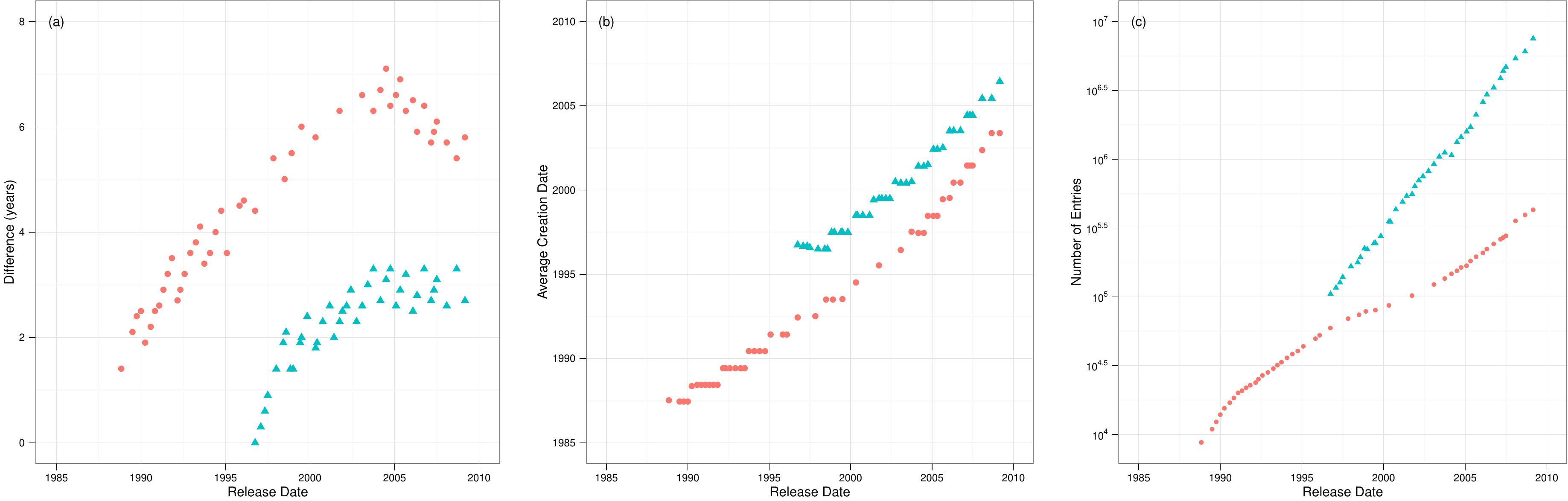}
 \caption{\swp (red circles) and \tr (blue triangles). (a) Growth
   (number of entries) in \swp and \tr over time. (b) Average creation
   date over time for \swp and \tr. (c) Difference between release
   date and average creation date (i.e. age) over time.  }
  \label{fig:averageEntries}
\end{figure*}

Our prior analysis has investigated annotation quality in bulk,
without analysing how individual records are maturing. If we consider
``maturity'' as a simple function of age, then we would expect, given
the rapid increase in the size of \swp, while new records appear, the
older entries should mature. Figure~\ref{fig:averageEntries}(a) illustrates
the exponential rate at which \swp and \tr are growing, showing the
number of entries in each database version.

Each entry contains a date stamp, indicating when it was first introduced into
the database. We use this information to show that the average creation date
of a record has increased only slowly over the life span of \swp as a whole --
illustrated in Figure~\ref{fig:averageEntries}(b). \swp is currently around 20
years old, yet the average record age is around 5 years old and decreasing.
This is illustrated in Figure~\ref{fig:averageEntries}(c), where we show the
difference between the average creation date and release date. As an example,
\swp version 9 was released in November 1988 and the average entry release
date is July 1987, so the figure reflects this difference of 1 year and 4
months. Given this, we wished to abstract from increasing size of \swp and ask
whether individual records appear to be maturing.

This analysis is not straight-forward; we need, essentially, a set of records
which relate to a defined set of proteins. To achieve this we extracted the
annotations from all of the entries that were common in both \swp version 9
and \up/\swp version 15 (the first and last version available to us). The
resulting $\alpha$ values are shown in Figure~\ref{fig:maturealphas}(b), with
the addition of the $\alpha$ value for those entries in \up version 15 but not
\swp version 9. These results show that the $\alpha$ value for the mature set
of entries has decreased over time, correlating with the \swp database as a
whole.

Given that the $\alpha$ value for mature entries has decreased over
time, it is of interest to investigate the $\alpha$ values of entries
that are new to each version of \swp. For this, we extracted
annotations from entries that appeared for the first time in a given
database version. Results of this analysis are shown in
Figure~\ref{fig:maturealphas}(a). It again would appear that the
$\alpha$ value is decreasing over time, similar to that of other \swp
graphs.

\section{Discussion}

The biological community lacks a generic quality metric that allows
biological annotation to be quantitatively assessed and compared. In
this paper we applied \pw distributions to the \up database and
linked the extracted $\alpha$ values to Zipf's principle of least
effort in an attempt to derive such a generic quality metric. The
results within this paper give confidence to our initial hypothesis
that this approach holds promise as a quality metric for textual
annotation.

Initially, our analysis focused on the manually-curated \swp. As shown
in Figure~\ref{fig:tremblvsswissprot_alphas}, early versions of \swp
give $\alpha$ values that suggest annotations were of high quality;
that is, they were of least effort for the reader. However, over time
we see a steady reduction in the $\alpha$ value, which does suggest
that the average annotation is now harder for readers to interpret,
requiring more expertise to consume the data than was previously
required. This result is perhaps best explained by the exponential
increase of data that is added to \swp. Manual annotations are
regarded as the highest quality annotation available and it is
inevitable that the acknowledged pressure on manual annotation is
going to increase. This conclusion appears to fit with previous
research~\citep{Manual07Baumgartner} that shows manual curation
techniques cannot keep up with the increasing rate of data.

Many of the patterns exhibited by \swp are also shown in our analysis
of \tr. From Figure~\ref{fig:tremblvsswissprot_alphas}, we conclude that
annotation in \swp and \tr show similar characteristics in that, for
both cases, annotation appears to be increasingly optimized to
minimize efforts for the annotator rather than the reader;
unsurprisingly, this appears to be more pronounced for \tr than for
\swp. We are currently unclear whether this form of direct
numerical comparison over the two different resources is highly
meaningful, although this distinction between the two resources appears to be
more pronounced over time rather than less. Therefore, these results
are consistent with the conclusion that manual annotation behaves as a
significantly more mature language than automated annotation. This
fits with our \textit{a priori} assumptions which again is suggestive
that this form of analysis is operating as a measure of quality. 

In addition to analysing whole \up datasets, we also investigated how
sets of entries mature over time (Figure~\ref{fig:maturealphas}(b))
and the quality of annotations within new entries
(Figure~\ref{fig:maturealphas}(a)). Within the mature entries we
interestingly see a decrease in quality over time, rather than
increasing or maintaining a similar quality level. However this
decrease is much slower than the \swp database as a whole over time,
and is still of much higher quality than the remainder of entries in
\up version 15. For the new annotations we also see a general decrease
in quality over time. It is plausible that these results stem from the
manner of management and curation of annotations; the UniProt
annotation protocol consists of six key
steps~\citep{Magrane11UniProt}, one of which is identifying similar
entries (from the same gene and homologs by using BLAST against \up,
UniRefs and phylogenomic resources). If two entries from the same gene
and species are identified then they are merged; annotations between
the remaining entries are then standardised. It would appear that
attempts to standardise growing sets of similar entries is having a
detrimental effect on the quality of both individual entries and the
overall database.

In addition to being used as a quality measure, the approach described
here could be used for artifactual error detection. Our early analysis
identified information with no biological significance (copyright
statements) included within the comment lines.

Our focus in this paper was on \up, and we have not tested across other
databases. One database of immediate interest would be InterPro. The work in
this paper focuses on protein annotation; extending this to InterPro would
allow us to analyse protein family annotation which would normalize for the
many near duplicate records of the large protein families found in \up. This
would require further bulk analysis -- however, once data is correctly
extracted, this form of analysis is straight-forward and does not require
specialist resources. Analysis of other forms of annotation would also be
interesting; \citeauthor{Kalankesh12Language}~(\citeyear{Kalankesh12Language})
has recently reported on similar results in Gene Ontology annotation.

Further work analysing additional databases would allow us to draw more
conclusive conclusions regarding the fitting of the \pw, and consequently the
usability of $\alpha$ as a quality metric. However, our analysis of \up
suggests that this approach holds promise of being a useful tool for database
curators and users alike.

\section*{Acknowledgements}

The authors thank Allyson Lister for her helpful
discussions. We also thank the UniProt helpdesk for answering our various
queries and Daniel Barrell at EBI for making available the historical versions
of \tr. This work was partly supported by EPSRC.

\bibliographystyle{natbib}
\bibliography{UniProt_Annotation_Bell_Gillespie_Swan_Lord}

\end{document}